%% file: main.tex
\documentclass[10pt,conference]{IEEEtran}
\IEEEoverridecommandlockouts

\usepackage{cite}
\usepackage{tikz}
\usepackage{url} 
\usepackage{rotating}       
\usepackage{graphicx}       
\usepackage{amsmath,amssymb,amsfonts} 
\usepackage{textcomp}

\usepackage{booktabs}
\usepackage{multirow}
\usepackage{array}
\usepackage[table]{xcolor}
\definecolor{lightgreen}{RGB}{144, 238, 144}

\usepackage{tcolorbox}
\usepackage{pifont}
\usepackage{footnote}
\usepackage{enumitem}
\usepackage[ruled,vlined]{algorithm2e}
\usepackage{listings}

\def\BibTeX{{\rm B\kern-.05em{\sc i\kern-.025em b}\kern-.08em
    T\kern-.1667em\lower.7ex\hbox{E}\kern-.125emX}}
\begin{document}

\title{Lares: LLM-driven Code Slice Semantic Search \\for Patch Presence Testing}

\author{\IEEEauthorblockN{Siyuan Li\textsuperscript{1,2,3}, Yaowen Zheng\textsuperscript{2,3}, Hong Li\textsuperscript{2,3,*}, Jingdong Guo\textsuperscript{2,3}, Chaopeng Dong\textsuperscript{2,3}\\
Chunpeng Yan\textsuperscript{2,3},  Weijie Wang\textsuperscript{2,3}, Yimo Ren\textsuperscript{2,3}, Limin Sun\textsuperscript{2,3}, Hongsong Zhu\textsuperscript{2,3}}
\IEEEauthorblockA{\textit{\textsuperscript{1}School of Cyber Science and Technology, Shandong University, Shandong, China}}
\textit{\textsuperscript{2}Institute of Information Engineering, Chinese Academy of Sciences, Beijing, China} \\
\textit{\textsuperscript{3}School of Cyber Security, University of Chinese Academy of Sciences, Beijing, China}\\
\\
siyuan@sdu.edu.cn,  \{zhengyaowen,lihong,guojingdong,dongchaopeng,yanchunpeng\}@iie.ac.cn,\\
\{wangweijie,renyimo,sunlimin,zhuhongsong\}@iie.ac.cn
 \thanks{* corresponding author: lihong@iie.ac.cn}}

\maketitle

\begin{abstract}
In modern software ecosystems, 1-day vulnerabilities pose significant security risks due to extensive code reuse. Identifying vulnerable functions in target binaries alone is insufficient; it is also crucial to determine whether these functions have been patched. Existing methods, however, suffer from limited usability and accuracy. They often depend on the compilation process to extract features, requiring substantial manual effort and failing for certain software. Moreover, they cannot reliably differentiate between code changes caused by patches or compilation variations.

To overcome these limitations, we propose Lares, a scalable and accurate method for patch presence testing. Lares introduces Code Slice Semantic Search, which directly extracts features from the patch source code and identifies semantically equivalent code slices in the pseudocode of the target binary. By eliminating the need for the compilation process, Lares improves usability, while leveraging large language models (LLMs) for code analysis and SMT solvers for logical reasoning to enhance accuracy. Experimental results show that Lares achieves superior precision, recall, and usability. Furthermore, it is the first work to evaluate patch presence testing across optimization levels, architectures, and compilers. The datasets and source code used in this article are available at https://github.com/Siyuan-Li201/Lares.
\end{abstract}


\input{tex/intro}
\input{tex/motivation}

\input{tex/approach}

\input{tex/implement}

\input{tex/eval}
\input{tex/discuss}
\input{tex/related}

\section{Conclusion}
It is crucial to determine whether the target function in binary has been patched. In this paper, we presented Lares, a novel method for patch presence testing. Unlike traditional methods that rely on compilation process to extract patch features, Lares directly leverages source code functions and patch information, providing a lightweight and scalable solution for cross-architecture detection. By employing LLM-driven semantic analysis, Lares accurately identifies patch code slices, significantly improving precision and recall. 
We compare Lares with existing methods in detail and discuss the contribution of each component of Lares. 
Our evaluation demonstrates that Lares not only achieves high scalability but also effectively handles diverse architectures and compilation environments. 

\section{Acknowledge}
This work was supported by Key R\&D Program of Shandong Province, China(No. 2024CXGC010114), National Natural Science Foundation of China under Grant(No. 62372268), Shandong Provincial Natural Science Foundation, China (No. ZR2022LZH013, No. ZR2021LZH007). Any opinions, findings and conclusions in this paper are those of the authors and do not necessarily reflect the views of the funding agencies.

\bibliographystyle{unsrt}
\bibliography{ref}

\end{document}

%% file: tex/intro.tex
\section{Introduction}

Software development is not a repetitive process. Developers often reuse third-party libraries to accelerate development, resulting in a complex software supply chain ecosystem \cite{FirmSec}. 
However, this practice introduces security risks by propagating vulnerabilities. 
According to the Synopsys report \cite{Blackduck}, 96\% of tested software reused at least one third-party library, with 89\% of codebases containing open-source components that had not been updated for over two years, and some for more than four years. The failure to update and patch vulnerabilities promptly exacerbates the impact of 1-day vulnerabilities.

\begin{figure}

\centering
    \includegraphics[width=\linewidth]{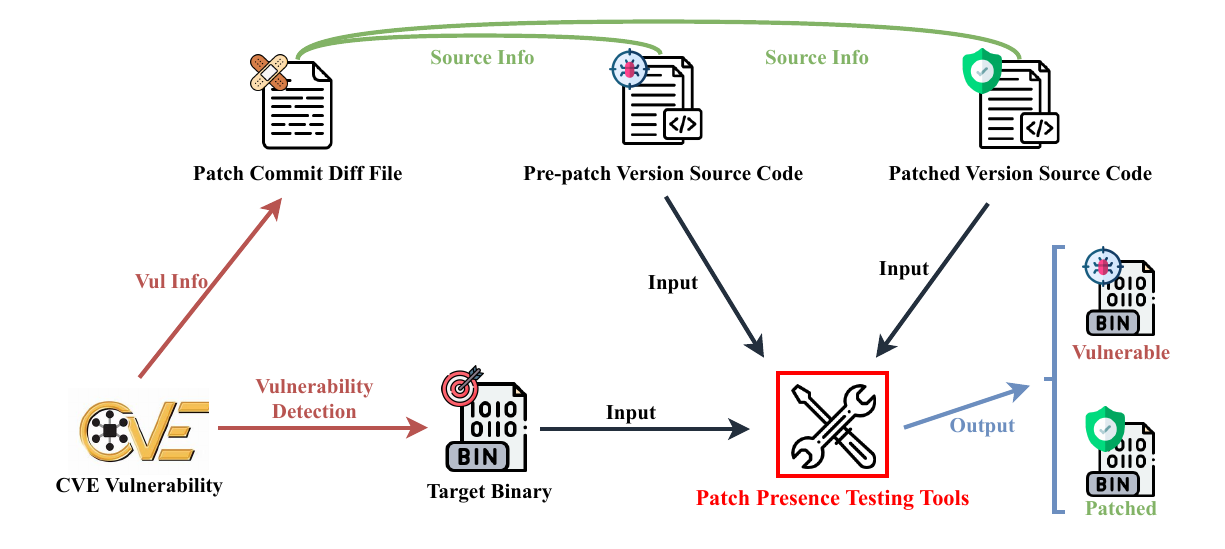}

\caption{The Background.}
\label{fig1}
\end{figure}

Several approaches \cite{Gemini, jTrans, Vulhawk, FirmSec, LibDB, LibAM, binaryai} have been proposed for detecting 1-day vulnerabilities in software. These approaches typically collect third-party libraries and vulnerable functions, performing function- or library-level matching using binary similarity techniques. However, they primarily identify functions similar to vulnerable ones, struggling to differentiate between vulnerable and patched functions due to minimal modifications in patches \cite{robin}. To address this limitation, patch presence testing was introduced, first proposed by Fiber in 2018 \cite{Fiber}. As in Figure~\ref{fig1}, this technique analyzes a target function and patch information (e.g., vulnerable and patched versions) to determine whether the target function is closer to the vulnerable or patched version, enabling fine-grained distinction. We classify patch presence testing methods into two categories: syntactic-based and semantic-based approaches.

Syntactic-based methods usually use syntactic features for detection. BinXray \cite{BinXray} uses the sequence of mnemonic operators in binary instructions, function calls and constant values for detection, and combines structural statistics such as the number of instructions, basic blocks,
branches and the control flow graph. Fiber \cite{Fiber} extracts the differences in control flow graphs and abstract syntax trees from the vulnerable version and patch version of the kernel to detect patches in the target kernel. PatchDiscovery \cite{patchdiscovery} uses basic blocks with more changed instructions as representative blocks to improve the performance of BinXray \cite{BinXray}.

Semantic-based methods usually use symbolic execution to extract various types of semantic features. PDiff \cite{Pdiff} first locates the anchor block in CFG, and then uses symbolic execution to obtain semantic features such as path constraints to reach the anchor block for detection. Robin \cite{robin} generates a malware function input (MFI) through symbolic execution from reference binaries, and collects four runtime features to compare whether the target binary is closer to the vulnerability version or the patch version. PS3 \cite{ps3} also uses symbolic simulation to extract four similar semantic features (write data to reg, Store data to mem, Condition, call\&Return) for comparison. 



However, it is essential to note that existing methods exhibit certain limitations, which hinder their performance and usability in the real world. We summarize the limitations of the existing works in the following three points.

Firstly, existing methods rely on compilation to extract features, which is often difficult to automate and may fail for many software projects, reducing usability (\textbf{P1}). Patch information resides at the source code level, while the detection target is a binary file. Current approaches compile both vulnerable and patched versions into binaries \cite{robin,ps3,BinXray} or intermediate representations (IR) \cite{react} to extract features using debugging information. 
As a result, these methods rely on manual compilation, making them unsuitable for large-scale deployment. To address these limitations, we propose to eliminate the need for compilation, directly extracts features from patches and source code, and enables efficient, automated large-scale detection.

Secondly, existing methods struggle to locate patch-related code fragments accurately, reducing detection accuracy (\textbf{P2}). Binary similarity matching typically operates at the function or library level, whereas patch detection requires finer-grained comparison of specific code fragments. While extracting patch-related fragments from reference binaries (e.g., vulnerable and patched functions) is straightforward, identifying corresponding fragments in target binaries remains challenging. Existing methods \cite{ps3, robin} tend to directly match all features of the entire function with patch features. An effective method is needed to precisely locate and extract the corresponding code fragments from target binaries.

Thirdly, existing methods struggle to maintain accuracy across different compilation environments (\textbf{P3}). In real-world scenarios, target binaries may be compiled using varying optimization options, compilers, or architectures, making it difficult to distinguish between changes introduced by patches and those caused by compilation differences. Existing approaches partially mitigate the impact of optimization options but still suffer significant accuracy degradation. Furthermore, most existing approaches \cite{BinXray, robin, ps3} only support feature comparison within the same architecture and do not support cross-architecture detection. In addition, they need to write feature extraction code specifically for each new architecture. 
Only REACT claims to support cross-architecture detection from the IR level, but it has not been evaluated on a cross-architecture dataset.



To solve the problems mentioned above, we propose Lares, an LLM-driven code slice semantic search technique that enhances the usability and accuracy of binary patch presence testing. Lares identifies the code slice corresponding to the patch source code from decompiled pseudocode and compares their semantic differences. Unlike existing methods, Lares does not require a compilation process and instead extracts features directly from the patch diff and patch function source code (\textbf{for P1}). It uses the decompiled pseudocode of the target binary function for comparison, with features that are architecture-independent and applicable to stripped binaries (\textbf{for P3}). Additionally, inspired by prior work on LLMs in decompilation \cite{llm4decompile} and function summarization \cite{degpt}, Lares employs LLMs to locate patch-related fragments, ensuring consistent granularity for precise comparison (\textbf{for P2}).

Lares comprises three modules. First, the patch enhancement module expands patch semantics by adding context-related statements to patch code with few statements. Second, the patch localization module leverages an LLM to identify pseudocode slices corresponding to patch code slices within the target function's pseudocode. Finally, the patch verification module validates the localization results to determine whether the target corresponds to a vulnerability or a patch by LLM and SMT Solver. The core insight of Lares lies in utilizing the code analysis and logical reasoning capabilities of LLMs to locate patch-related statements in pseudocode and compare them with source code slices from both the pre-patch and patched versions.

We thoroughly analyze the issues introduced by the compilation process in existing methods and demonstrate how Lares naturally avoids them. Our evaluation shows that Lares's lightweight design not only maintains usability but also improves precision and recall by effectively addressing P2 and P3. Compared to existing methods, Lares achieves a significant F1 score improvement of 9\%-10\%. This balance of accuracy and usability makes Lares a more effective solution for patch presence testing.

We summarize our main contributions below:

\begin{itemize}

\begin{figure*}

\centering
    \includegraphics[width=\linewidth]{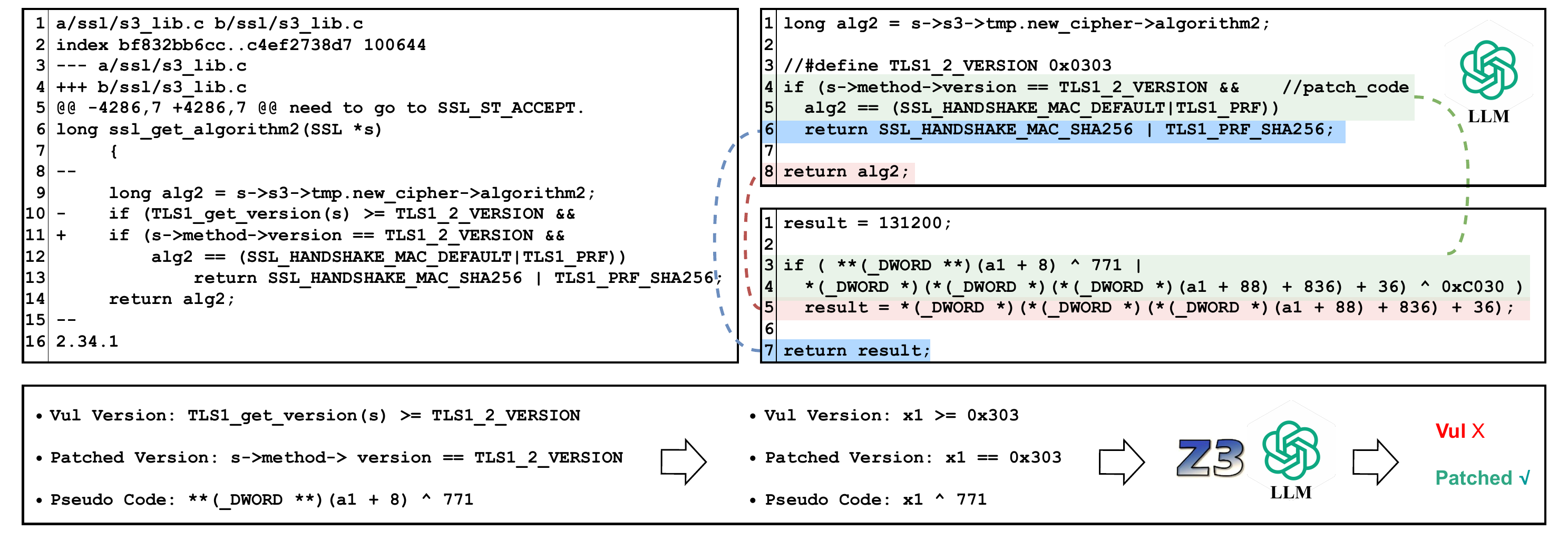}

\caption{A motivation example.}
\label{fig:motivation}
\vspace{-1.0em}
\end{figure*}




\item We propose Lares, the first compile-free, LLM-based approach for patch presence testing tasks, which eliminates substantial manual work and achieves large-scale automated detection.
\item We implement Lares with 3,000 lines of Python code and integrate it with several source and binary analysis tools, while also validating its performance using advanced LLMs.
\item We conduct an extensive evaluation of Lares, and the results show that Lares outperforms state-of-the-art approaches in terms of both accuracy and recall overall across a range of architectures and compilation options.

\end{itemize}


%% file: tex/motivation.tex
\section{Motivation}

We illustrate our motivation and insights with the example shown in~Figure~\ref{fig:motivation}. First, we demonstrate the necessity of conducting patch presence testing. Next, we analyze the limitations of existing approaches and present our insights for developing more robust tools for patch presence testing.

\subsection{Necessity of Patch Presence Testing}
Figure~\ref{fig:motivation} illustrates CVE-2013-6449 as a representative example. The first box on the left contains the patch commit for the vulnerability, which introduces minimal code changes. Specifically, it replaces a version-related macro definition with a structure pointer and modifies the comparison operator from ``\textgreater=" to ``==". 
These subtle changes render the patched function highly similar to the vulnerable one. Consequently, existing binary similarity detection methods struggle to differentiate between the vulnerable and patched versions, as they primarily identify function-level similarities. This highlights the need for a more fine-grained patch existence verification method to detect the presence of the vulnerability accurately.

\subsection{Limitations and Insights}
\subsubsection{Reliance on Project Compilation}

Existing methods rely heavily on the compilation process to extract features, such as control flow graphs (CFG) \cite{BinXray}, assembly code \cite{robin, ps3}, or intermediate representations (IR) \cite{react}. For each vulnerability to be detected, these methods require manually compiling both pre-patch and patched binaries. This process necessitates the inclusion of debugging information to identify the precise location of the patch code within the binary. While B2SFinder \cite{B2SFinder} attempted to automate the compilation process, they reported that only 25\% of the GitHub projects they tested could be successfully compiled automatically. Furthermore, we observed that compilation is not only challenging to fully automate but often requires extensive domain knowledge, with many projects failing even under manual compilation. These limitations render existing methods insufficient for real-world applications. Table~\ref{tab:usability} describes the specific problems caused by the compilation process.




\begin{table}
  \centering
  \caption{Usability Evaluation of Existing Methods}
  \vspace{0.5em}
  \label{tab:usability}
  \setlength{\tabcolsep}{4pt}
  \begin{tabular}{cccccc}
    \toprule
     Problems & REACT & BinXray & Robin &  PS3 & Lares\\
    \midrule
    ED & \ding{56} &  \ding{56} & \ding{56} & \ding{56} & \ding{52}\\
    CP & \ding{56} &  \ding{56} & \ding{56} & \ding{56} & \ding{52}\\
    CV & \ding{56} & \ding{56} &  \ding{56} & \ding{56} & \ding{52}\\
    BT & \ding{56} & \ding{56} &  \ding{56} & \ding{56} & \ding{52}\\
    AC & \ding{56} & \ding{52} &  \ding{52} & \ding{52} & \ding{52}\\
  \bottomrule
  \vspace{-1.5em}
\end{tabular}

\smallskip
\noindent\small 
\textit{\textbf{Note:}} ED: External Dependencies. CP: Compilation Parameters. CV: Compiler Versions. BT: Build Tools. AC: Additional Compilation. \ding{52} indicates that this method can solve the problem, while \ding{56} indicates that it cannot.

\end{table}

\textbf{External Dependencies:} Some projects require resolving external library dependencies during compilation (e.g., Zlib for OpenSSL, Libpcap for Tcpdump), but they do not provide an automatic way to build the dependencies. This demands domain knowledge and careful alignment of compilation parameters (e.g., the path of external libraries compiled manually), hindering automation and increasing the likelihood of failures in complex projects.

\textbf{Compilation Parameters:} Reference binaries must be compiled with specific parameters (e.g., -O0, -g). Manually configuring compiler and linker options in the project-specific config file further reduces usability.

\textbf{Compiler Versions:} Variations in compiler versions can cause compatibility issues. Older projects may fail with newer compilers due to stricter checks, while newer projects may rely on C++20 features supported only in g++ v10 or later.

\textbf{Build Tools:} Projects use diverse build tools (e.g., Automake, CMake, Bazel), which complicates the setup process. Datasets from prior work \cite{ISRD} such as exjson even require MIX as the build tool.

\textbf{Additional Compilation:} Methods like REACT require extracting the compilation parameters during the compilation process and re-compiling to generate intermediate representations (IR) using Clang, further complicating the process and increasing the risk of errors.


Therefore, we aim to propose a \textbf{compile-free approach} for patch presence testing. Unlike existing methods, we decompile the target binary into pseudocode and analyze it by extracting features from the source code for comparison to determine the presence of a patch. 
This compile-free design inherently addresses all issues in the compilation process.

\begin{figure*}
\centering
    \makebox[\textwidth][c]{%
        \includegraphics[width=1.05\textwidth]{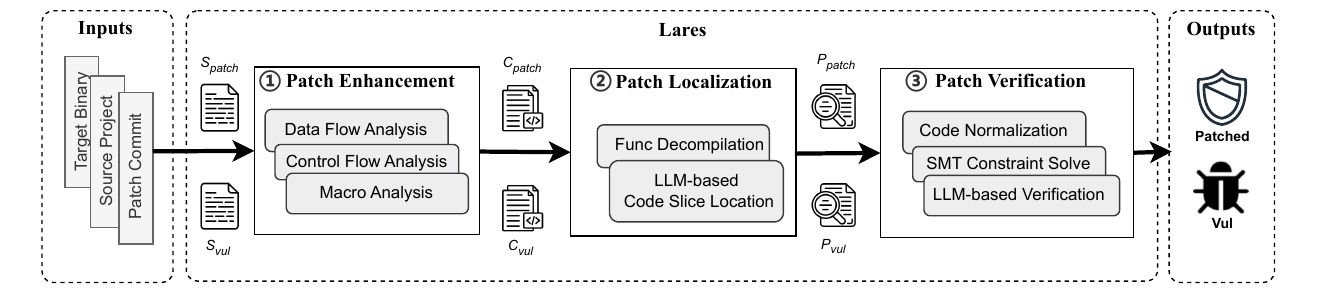} 
    }

\caption{The workflow of Lares.}
\label{fig:workflow}
\vspace{-1.0em}
\end{figure*}

Avoiding compilation is a critical advantage. In practical application scenarios, the analysis targets are often real-world binaries extracted from software or firmware. Although these projects may be technically compilable, the compilation process is difficult to automate and often requires manual, case-by-case configuration. For large-scale vulnerability analysis tasks that demand efficient processing across numerous projects, these issues significantly hinder system efficiency and scalability.



\subsubsection{Inaccuracy in Patch Localization}

In addition, existing methods, such as PS3 \cite{ps3}, extract features from the entire function code in the target binary and match these features with those of vulnerabilities or patches. However, this approach introduces numerous irrelevant features, as patch statements often constitute only a small portion of the function. Other methods, including BinXray \cite{BinXray}, Robin \cite{robin}, and PDiff \cite{Pdiff}, rely on control flow graph (CFG) blocks and heuristic rules to locate anchor points. However, CFG blocks are sensitive to changes in the compilation environment, leading to inaccuracies in these methods. To address the performance degradation caused by inconsistent comparison units, we aim to locate the code slices in the target function that correspond to the patch code slices. This method should be robust to compilation environment variations and leverage semantic analysis to identify the relevant code slices for comparison.

Therefore, we aim to propose an approach that can \textbf{accurately locate patch-related code within the target} and perform subsequent analysis and detection.
As shown in the two boxes in the upper corner of Figure~\ref{fig:motivation}, we directly compare the patch-related source code with the pseudocode obtained through decompilation of the target binary. Specifically, we leverage LLM to find pseudocode slices corresponding to patch-related source code from pseudocode, as highlighted in Figure~\ref{fig:motivation}. The dotted lines illustrate the correspondence between these statements. This patch localization enables comparison at a consistent and fine-grained code slice level.

However, LLMs cannot perfectly locate the patch or vulnerability slice. While LLMs can identify the pseudocode slice most similar to the target patch fragment, the retrieved slice may precede or follow the actual patch. Therefore, an additional patch verification step is necessary. Similar to prior work \cite{robin, ps3, react}, we employ an SMT solver for semantic-level comparison. As shown in the bottom box of Figure~\ref{fig:motivation}, we extract key instructions from the code fragment and normalize them into unified equations. Using the Z3 solver, we evaluate semantic equivalence, effectively handling cases such as $x1 \mathbin{\hat{}} 771$ and x1 == 0x303, which differ syntactically but are semantically equivalent. For cases where the Z3 solver fails, we leverage LLMs to further analyze the match between the target code slice and the vulnerability or patch slice to determine the correct correspondence.


%% file: tex/approach.tex
\section{Methodology}


In this section, we introduce the design of Lares. We begin by explaining its overall workflow, followed by a detailed description of each module.

\subsection{Overview}

The workflow of Lares, illustrated in Figure~\ref{fig:workflow}, consists of three phases: patch enhancement, patch localization, and patch verification. We assume that a binary code similarity detection scheme (e.g., BinaryAI \cite{binaryai}) has been used to identify potential vulnerable functions $F_{t}$ in the target binary, and the goal of this work is to further verify whether the vulnerability has been patched. 
We take the target binary, source code project, and patch commit as input. Then, Lares determines whether the vulnerable code in the target binary has been patched.

\textbf{Patch Enhancement}. 
Given the source code of the vulnerable function $S_{vul}$ and that of the patched function $S_{patch}$ as input, this module aims to extract patch-related code slices for both versions ($C_{vul}$ and $C_{patch}$) to allow subsequent patch localization within the target. Specially, some patches modify only a small portion of the source code, sometimes as little as a single line, making the code diff between the vulnerable and patched versions insufficient. Therefore, the goal is to extract more enriched code slices that include only statements relevant to the patch and avoid the introduction of irrelevant code.


\textbf{Patch Localization}. 
In this step, we first decompile $F_{t}$ into pesudocode $P_{t}$.
Next, we aim to locate the corresponding slice within $P_{t}$ using $C_{vul}$ and $C_{patch}$ obtained from the previous step. 

\begin{figure}

\centering
    \includegraphics[width=\linewidth]{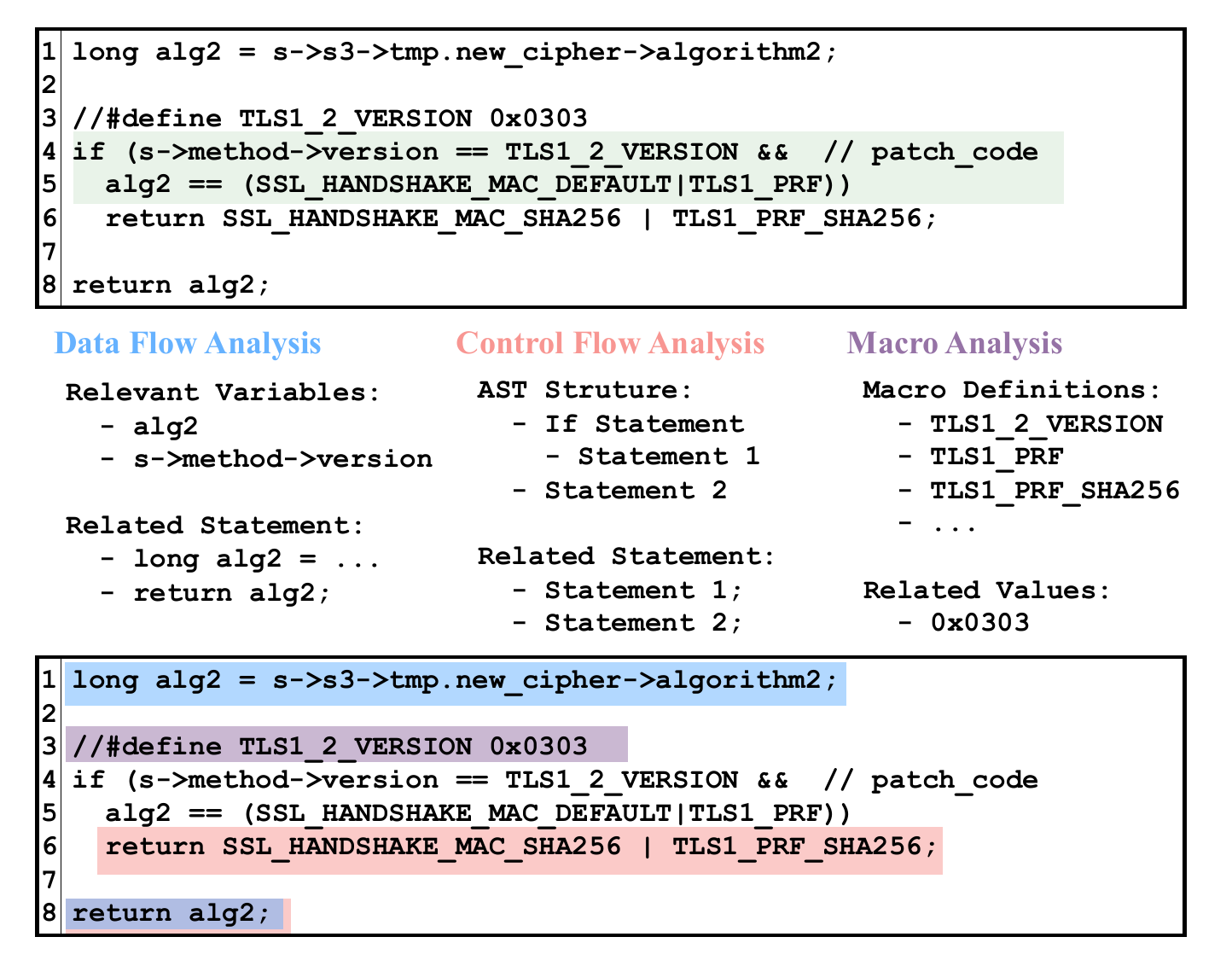}

\caption{The Patch Enhancement Module.}
\label{fig:patch_enhance}
\vspace{-1.0em}
\end{figure}

Currently, LLMs have demonstrated strong code analysis capabilities in tasks such as code decompilation \cite{llm4decompile} and function summarization \cite{degpt}. 
Therefore, we designed an LLM-based approach to locate code slices within $P_{t}$. 
Specifically, we crafted a tailored prompt containing $P_{t}$, $C_{vul}$ and $C_{patch}$. The LLM is tasked with mapping each line of the source code slice to its corresponding pseudocode sequence, and finally get the pseudocode slices ($P_{vul}$ and $P_{patch}$) corresponding to $C_{vul}$ and $C_{patch}$ .
While this result identifies the pseudocode slices most semantically similar to the source code slice, further verification is required to determine whether the fragment represents the vulnerability or the patch.

\textbf{Patch Verification}. 
From the previous step, we obtained two pairs, ($C_{vul}$, $P_{vul}$) and ($C_{patch}$, $P_{patch}$), representing the source code slices before and after the patch, along with their potential corresponding pseudocode slices in the target. In this step, we analyze the semantic equivalence of each pair. The pair with the highest semantic equivalence similarity is considered the best match, allowing us to infer whether the target corresponds to the pre-patch or patched version.


    
    


\subsection{Patch Enhancement}


In this phase, we utilize three core strategies to enrich patch-related code slices, including data flow analysis, control flow Analysis, and macro analysis.
Given a patch commit diff file and the pre-patch and patched versions of the function source code ($S_{vul}$ and $S_{patch}$), the module outputs the patch-related code slices for both versions ($C_{vul}$ and $C_{patch}$).  Similar to existing work MVP \cite{MVP}, Lares leverages lightweight slicing implemented via Joern \cite{joern}, enabling a compile-free patch presence test. This method is intra-function.

\subsubsection{Data Flow Analysis}

The first enhancement strategy targets the data dependencies of the patch. Variables defined or used within the patch often act as critical links between the patch and the rest of the program. However, since patches typically modify only a small portion of the code, the broader context of these variables often remains unexplored. To address this, we conduct systematic data flow analysis to capture the complete role of these variables within the program.

We begin by parsing the patch code to extract all defined variables and used variables. These variables represent key points of interaction between the patch and its surrounding code. For instance, as shown in Figure~\ref{fig:patch_enhance}, a patch modifying a single \texttt{if} statement would involve the condition of the \texttt{if} clause and any variables referenced within its body, which may significantly interact with other parts of the program.

After identifying the relevant variables, we analyze the broader function or module containing the patch to locate additional statements that define or use these variables. For example, in Figure 4, lines 1 and 8 are data flow-related statements connected to the patch statement, providing additional context for analysis.


\subsubsection{Control Flow Analysis}

While data flow analysis focuses on variable interactions, control flow analysis examines the structural and logical organization of the code impacted by the patch. This analysis is crucial for understanding how the patch affects the program’s execution paths, particularly in the presence of complex control structures such as loops and conditional branches. By incorporating these dependencies, control flow analysis offers a comprehensive understanding of the patch's impact on program execution.


For each patch, we determine the control flow structure in which it resides. This involves identifying the entry points of control structures, such as the condition in an \texttt{if} statement or the header of a loop. We further analyze the structure by extracting the first statement of each block and the statement immediately following the block. For example, if a patch modifies the body of an \texttt{if} statement, we also capture the corresponding \texttt{else} branch,  if present, along with the statement following the \texttt{if-else} block.

\subsubsection{Macro Analysis}

In addition to data flow and control flow, the third enhancement strategy focuses on macro definitions, which are commonly used in source code to represent constants, expressions, or inline code fragments. Since the target binary is stripped of such definitions, we must identify and resolve the specific values of macro definitions in the patch to better align with the values found in the pseudocode.


After identifying the relevant macro definitions, we locate their corresponding values or code fragments in the source code. These values are then substituted into the patch, replacing macro placeholders with their concrete representations. For example, in Figure~\ref{fig:patch_enhance}, the macro \texttt{TLS1\_2\_VERSION} resolves to the value \texttt{0x0303}, enabling correct matching in later stages (e.g., aligning 0x303 with 771 via macros). Although the direct search strategy carries some risk of error, this lightweight approach is effective in practice and requires no compilation.

\subsection{Patch Localization}

\begin{table*}
  \centering
  \caption{Prompt for each task in Lares}
  \label{tab:freq111}
  \vspace{0.5em}
  \renewcommand{\arraystretch}{1.5}  
  \begin{tabular}{|m{0.27\textwidth}|m{0.78\textwidth}|}
    \cline{1-2}
    \multicolumn{1}{|c|}{\textbf{Task}} & \multicolumn{1}{c|}{\textbf{Prompt Template}} \\
    \cline{1-2}
    \multicolumn{1}{|c|}{Patch Localization} & Suppose you are a software reverse engineer with strong code analysis skills. You have the source code of a function and the pseudo code obtained through binary decompilation. Lines in the source code that end with ``//patch\_code" are patch codes. Can you identify the patch codes in the pseudo code that corresponds to the patch code? Must only output your findings as a JSON dictionary.
    
    - Output format:\textless json\_format\_sample\textgreater 
    
    - Source code: \textless source\_code\textgreater
    
    - Pseudocode: \textless pseudo\_code\textgreater\\
    \cline{1-2}
    \multicolumn{1}{|c|}{Patch Verification} & You are a software reverse engineer analyzing decompiled pseudo code. Your task is to determine whether the code is patched or pre-patch version by analyzing the reliability of matching results. Must only output your findings as a JSON dictionary.

- INPUT:
1. Diff File: \textless patch\_diff\_label\textgreater
2. patched version matches: \textless patch\_result\_json\textgreater
3. pre-patch version matches: \textless vul\_result\_json\textgreater

- ANALYSIS REQUIREMENTS: Evaluate each match in patched and pre-patch: Semantic correctness, Logic consistency, Context compatibility, Potential false matches. Compare quality of matches: Which version has more reliable matches, Which matches might be incorrect, Overall semantic alignment

- RULES: Only one result (patched version or pre-patch version) corresponds to the correct version. Better semantic match determines the version
 \\
    \cline{1-2}
  \end{tabular}
\end{table*}

The purpose of the Patch Localization module is to identify the corresponding pseudocode slice ($P_{vul}$ and $P_{patch}$) within the target function's pseudocode, using the enhanced patch code slice from the previous module ($C_{vul}$ and $C_{patch}$). Rather than using only the patch code slice as input, we utilize the entire function's source code. To assist identification, we annotate each line of the patch code slice within the function source code by adding the comment ``\texttt{//patch\_code}".

\subsubsection{Truncation of Function}

To address the token limits of large language models (LLMs), we apply AST-guided truncation to retain essential contextual information while ensuring syntactic and semantic integrity. 

For patch code. we use the first statement of the patch as an anchor, and expand upward and downward to include surrounding lines of code. To ensure semantic integrity, truncation occurs at structural boundaries in the AST. For example, if the patch modifies a statement within a loop, the entire loop body is included. This process continues until the truncated code reaches a predefined token limit. The final patch code is annotated with the label ``\texttt{patch code}" for subsequent processing. We adopt a similar strategy for the pseudocode of the target function. Rather than extracting a single statement and its context, we extract multiple code slices truncated at the AST structure boundaries and input them into Lares for subsequent processing.

\subsubsection{Prompt Construction}

After truncating both the patch code and pseudocode, we construct prompts for the LLM to perform the patch localization task. The prompt follows a carefully designed template, as shown in Table~\ref{tab:freq111}. This template includes placeholders for the patch code and a single pseudocode segment, which are replaced with the truncated patch code and one segment of pseudocode, respectively. The prompt instructs the LLM to identify the pseudocode slice corresponding to the patch code slice. 

The constructed prompt is sent to the LLM, which analyzes the patch code and pseudocode segment and returns a JSON structure specifying the location of the patch code slice within the pseudocode. This JSON output includes the matched pseudocode slice and its corresponding location. The result is a precise mapping between the enhanced patch code slice and the corresponding pseudocode slice, providing the foundation for the subsequent \textit{patch verification} module.

\subsubsection{Reverse-Matching Strategy for Add and Delete Patches}

Not all vulnerabilities identified during the \textit{patch localization} module result in two matching pseudocode slices. For edit-type patches, it is possible to match the unique code before and after the patch with the target pseudocode, yielding two distinct results for the following verification. However, for add-only and delete-only patches, there is typically a single match between the added/deleted code slice and the pseudocode. This makes it challenging to determine whether the resulting pseudocode slice aligns more closely with the pre-patch or post-patch version. To address this, we introduce a novel \textbf{reverse-matching strategy} to ensure two matching results are always obtained.

Generating two pseudocode slices is practical and effective. Our goal is to compare the pseudocode with the two source codes before and after the patch. Because the patch-induced differences are minimal, an LLM cannot reliably locate a single pre- or post-patch slice. Instead, for each vulnerability, Lares matches the most similar pseudocode slice to the pre-patch source and another to the post-patch source, then selects the correct pairing to ensure accurate localization.

For add-only patches, in addition to the $P_{patch}$ that matches the added code slice ($C_{patch}$), the second matching result $P_{vul}$ is generated by reverse matching the pseudocode slice $P_{patch}$ with the source code of the pre-patch version $S_{vul}$. This determines whether the pseudocode slice originates from the newly added code or from existing lines in the pre-patch version. Similarly, for delete-only patches, the $P_{vul}$ is matched with the $S_{patch}$ to get the $P_{patch}$ for verification.

\subsection{Patch Verification}

The purpose of the patch verification module is to determine whether the pseudocode slice ($P_{vul}$ and $P_{patch}$) identified by the patch localization module truly corresponds to the vulnerable code or the patched code ($C_{vul}$ and $C_{patch}$). While the patch localization module identifies the pseudocode slice most similar to the patch code, additional analysis is required to distinguish whether the match aligns with the vulnerable or patched version. The Patch Verification module takes as input two locating results for the target function: one from the pre-patch version and one from the patched version. Its output is a determination of which locating result is correct.

Existing methods tend to use the Z3 solver for equivalence judgment. However, since feature extraction is performed on source code and pseudocode, it is difficult to fully obtain semantic formulas, which sometimes causes Z3 to fail. Therefore, we combined LLM with Z3 for verification. 
This verification process is fully automated and forms the final step in the pipeline, enabling accurate and efficient analysis of software patches.

\subsubsection{Lexical Analysis and Normalization}

In order to extract semantic formulas from code slices and perform SMT-Based constraint solving, we need to first normalize them. 
The process begins with a customized \textbf{lexical analysis} of the matched code slices, where the code is parsed into fundamental lexical units. Finally, we extract the normalized equation from the vocabulary unit.





        



We use a custom finite state machine (FSM) for lexical analysis. We take each line of code as input, provide it to the FSM character by character, and perform state transition according to each new character. This recursive definition ensures that the FSM processes the input string sequentially, one character at a time, maintaining a systematic and deterministic approach to lexical analysis.

After lexical analysis, we get the lexical units corresponding to the code. From these units, we extract four primary types of code slices:

\begin{itemize}
    \item \textbf{Conditional Statements:} e.g., \texttt{if (x > 0)}, \texttt{while (y == 1)}.
    \item \textbf{Assignment Statements:} e.g., \texttt{x = a + b}, \texttt{y = 0}.
    \item \textbf{Return Statements:} e.g., \texttt{return z}.
    \item \textbf{Function Calls:} e.g., \texttt{foo(x, y)}, \texttt{bar(x+14)}.
\end{itemize}

Next, all variables and complex operations on variables (e.g., pointer manipulations, structure operations, or unrecognized complex expressions) are normalized into \textbf{macro variables}. For example, an expression like \texttt{s->method->version == 0x303} is normalized into \texttt{x1 == 0x303}. This normalization ensures that the code slices are abstracted into a form suitable for semantic analysis.

\subsubsection{SMT-Based Constraint Solving}

Using the normalized code slices, we extract logical equations and expressions, which are then processed using the Z3 solver to check for semantic equivalence. To avoid interference from trivial or repetitive equations (e.g., \texttt{x1 == null}, \texttt{x1 == 0}), we extract the formulas that do not exist in the pre-patch version of the function as the unique features of the patched version. The same operation is also applied to the unique features of the pre-patch version. Only unique and meaningful equations are used for verification. If the Z3 solver finds semantically equivalent equations between the two slices, the verification is marked as successful.

\subsubsection{LLM-Based Semantic Reasoning}

\begin{table}
  \renewcommand{\arraystretch}{1.2} 
  \centering
  \caption{Dataset}
  \vspace{0.5em}
  \label{tab:dataset}
  \begin{tabular}{cccccc}
    \toprule
     Project & CVE & $func_v$ & Version & Binary & Testcase\\
    \midrule
    OpenSSL & 34 &  100 & 20 & 400 & 2000\\
    Freetype & 7 &  18 &  5 & 100 & 360\\
    Tcpdump & 25 &  78 &  2 & 40 & 1560\\
    Libxml2 & 7 &  28 &  4 & 80 & 560\\
    \midrule
    All & 73 &  224 & 31 & 620 & 4480\\
  \bottomrule
\end{tabular}

\end{table}

For cases where SMT-based constraint solving cannot directly verify equivalence (e.g., lack of unique logical structures), we employ a Large Language Model (LLM) for semantic reasoning. The LLM analyzes the two matching results (pseudo code vs. vulnerable source code slice and pseudo code vs. patched source code slice) to infer semantic equivalence. The LLM is tasked with identifying whether the pseudo code slice represents the vulnerable functionality or the patched functionality. We constructed Prompt as shown in Table~\ref{tab:freq111}. This reasoning step leverages the LLM's ability to understand high-level semantics and abstract relationships, providing a fallback mechanism for cases that are too complex for SMT solvers alone.

%% file: tex/implement.tex
\section{Implementation}

Lares is implemented in approximately 3,000 lines of Python code, integrating several tools for its functionality.

For binary analysis, we use IDA Pro \cite{IDA_Pro} to decompile binaries and extract function-level pseudocode via custom scripts. IDA Pro was selected for its robust decompilation and scripting capabilities. For source code, Tree-Sitter \cite{treesitter} parses code into functions with AST-based segmentation, while Joern \cite{joern} performs advanced static analysis, including data flow and control flow, using its code property graph model.

Semantic equivalence is determined with the Z3 SMT solver \cite{z3}, which verifies logical equivalences by normalizing code slices into macro-variables. For cases where formal methods are insufficient, Lares uses the Claude-3.5-Sonet \cite{claude} language model for semantic reasoning. Due to cost considerations, larger models like GPT-o1 were avoided, with vulnerability verification costing approximately \$0.20 per case. We use the default values for LLM parameters (temperature=1.0).

%% file: tex/eval.tex
\section{Evaluation}
We evaluate the effectiveness of Lares by answering the following research questions.



\textbf{RQ1}: How does Lares perform in the cross-optimization patch presence testing task compared to existing methods?

\textbf{RQ2}: How does Lares perform in the cross-architecture patch presence testing task? 

\textbf{RQ3}: How does every component in Lares affect the overall performance?

\textbf{RQ4}: How efficient is Lares? 

\paragraph{Experiment setup.}
The experiments are conducted on Ubuntu 22.04, powered by an Intel Xeon CPU with 128 cores at 3.0GHz and hyperthreading capabilities. 

\begin{table*}[t]
\renewcommand{\arraystretch}{1} 
\centering
\caption{Performance of patch presence testing with different optimization options on x86.}
\vspace{0.5em}
\label{table1}
\begin{tabular}{c|ccc|ccc|ccc|ccc|ccc|ccc}
\toprule[2pt]
\multicolumn{1}{c|}{\multirow{2}{*}{Model}} & \multicolumn{3}{c|}{O0}&\multicolumn{3}{c|}{O1}&\multicolumn{3}{c|}{O2}&\multicolumn{3}{c|}{O3}&\multicolumn{3}{c|}{Average}&\multicolumn{3}{c}{$Average_{new\_vul}$}\\
\cline{2-19}

\multicolumn{1}{c|}{} & P & R & F1 & P & R & F1 & P & R & F1 & P & R & F1 & P & R & F1 & P & R & F1\\
\hline


\multirow{1}{*}{BinXray} &1.0	&0.49	&0.66	&-	&-	&- &-	&-	&- &-	&-	&-	&-	&- &- &-	&- &-\\

\multirow{1}{*}{Robin} &0.64	&0.75	&0.69	&0.63	&0.70	&0.67 &0.62	&0.69	&0.66 &0.63	&0.73	&0.68 &0.63	&0.72	&0.67 & 0.69 & 0.73 & 0.70\\

\multirow{1}{*}{PS3} &0.76	&0.93	&0.83	&0.56	&0.70	&0.62 &0.56	&0.74	&0.64 &0.56	&0.74 &0.64 &0.61	 &0.78	&0.68 & 0.62 & 0.79 & 0.69\\

\multirow{1}{*}{Lares} &0.69	&0.91	&0.79	&0.71	&0.77	&0.74 &0.72	&0.79	&0.75 &0.78	&0.83	&0.81 &\textbf{0.72}	&\textbf{0.83}	&\textbf{0.77} & 0.76 & 0.81 & 0.79\\

\bottomrule[2pt]
\end{tabular}
\label{table_MAP}
\end{table*}

\begin{table*}[t]
\renewcommand{\arraystretch}{1} 
\centering
\caption{Patch presence testing accuracy of different compilation enviroment.}
\vspace{0.5em}
\label{table2}
\begin{tabular}{c|ccc|ccc|ccc|ccc|ccc|ccc}
\toprule[2pt]
\multicolumn{1}{c|}{\multirow{2}{*}{Enviroment}} & \multicolumn{3}{c|}{O0}&\multicolumn{3}{c|}{O1}&\multicolumn{3}{c|}{O2}&\multicolumn{3}{c|}{O3}&\multicolumn{3}{c|}{Average}&\multicolumn{3}{c}{$Average_{new\_vul}$}\\
\cline{2-19}

\multicolumn{1}{c|}{} & P & R & F1 & P & R & F1 & P & R & F1 & P & R & F1 & P & R & F1 & P & R & F1\\
\hline


\multirow{1}{*}{x86-gcc} &0.66  &0.85 &0.75	&0.71  &0.79 &0.74 &0.68	&0.77	&0.72 &0.73	&0.79	&0.76 &0.69	&0.80	&0.74 &0.74	&0.79	&0.76\\

\multirow{1}{*}{x64-gcc} &0.65	&0.85	&0.74	&0.69	&0.79	&0.73 &0.66	&0.79	&0.72 &0.73	&0.81	&0.76  &0.68	&0.81	&0.74   &0.74	&0.80	&0.77\\

\multirow{1}{*}{arm-gcc} &0.62	&0.80	&0.70	&0.64	&0.75	&0.69 &0.69	&0.75	&0.72 &0.70	&0.76	&0.73 &0.66	&0.76	&0.71  &0.68	&0.74	&0.71\\

\multirow{1}{*}{x86-clang} &0.69	&0.91	&0.79	&0.71	&0.77	&0.74 &0.72	&0.79	&0.75 &0.78	&0.83	&0.81 &0.72	&0.83	&0.77 & 0.76 & 0.81 & 0.79\\

\multirow{1}{*}{x64-clang} &0.68	&0.91	&0.78	&0.70	&0.77	&0.73	&0.71 &0.76	&0.73	&0.73  &0.79	&0.76	&0.71 &0.81	& 0.76 & 0.74 & 0.81 & 0.77\\


\bottomrule[2pt]
\end{tabular}
\label{table_MAP2}
\end{table*}

\paragraph{Evaluation metrics.}
We follow the evaluation criteria established in existing works~\cite{robin, ps3, react}, and use precision, recall and F1 score to evaluate the performance of Lares and baselines.
Specifically, $TP$, $FP$, and $FN$ refer to the number of patched functions truly classified as patched, vulnerability functions erroneously classified as patched, and patched functions erroneously classified as vulnerability, respectively. 






\paragraph{Dataset}
We select four real-world well-known projects from various application aspects for evaluation, which is the same as the existing works \cite{ps3, react}. These four projects involved protocol encryption, packet, XML analyzer, and font rendering. We compile the vulnerable and patched versions of OSS for each vulnerability collected using compiler gcc v9.4.0 and clang v6 with different optimization levels (O0 to O3) and different architectures (x86, x64, and ARM), respectively.
Table~\ref{tab:dataset} shows the statistical information for these projects. 
In total, we have obtained 73 CVEs with 224 distinct vulnerable and patched functions, denoted as the source functions. Finally, we conduct 20 * (112 + 112) = 4480 test cases.

In addition, we expanded the dataset to include 20 vulnerabilities disclosed after October 2024, totaling 400 cases, to ensure that they were not included in the training set of the model (claude-3.5-sonnet-20241022) to avoid data contamination.

\paragraph{Baselines}

We select three baselines in our experiments.


\begin{itemize}
    \item \textbf{BinXray \cite{BinXray}}: A classic syntactic-based approach for patch presence testing, widely adopted in various works.
    \item \textbf{Robin \cite{robin}}: A state-of-the-art semantic-based patch presence testing approach that leverages symbolic execution to extract patch features.
    \item \textbf{PS3 \cite{ps3}}: A state-of-the-art patch presence testing approach specifically designed to address challenges posed by compiler optimization.
\end{itemize}



Additionally, REACT \cite{react} needs to be compiled with clang to generate Intermediate Representations (IR). This process requires an additional compilation process after the normal compilation, which is more difficult to use than other methods. Since we aim to implement a compile-free method, REACT is orthogonal to our scope of consideration.



\subsection{Cross-optimization Task (RQ1)}

To evaluate Lares' performance in the cross-optimization (O0, O1, O2, and O3) patch presence testing task, we compare it against three state-of-the-art methods. The results are summarized in Table~\ref{table_MAP}.


BinXray \cite{BinXray} exhibits high precision under O0 (1.0). H
owever, it fails entirely under higher optimization levels (O1, O2, O3). This is because BinXray \cite{BinXray} heavily relies on syntactic features. Robin \cite{robin} demonstrates relative stability across different optimization levels, maintaining an F1 score in the range of 0.66 to 0.69. However, its overall accuracy and recall remain low compared to Lares. This is primarily because Robin \cite{robin} relies on symbolic execution for feature extraction, which fails for certain functions, and some extracted features are insufficient to distinguish between patched and unpatched code. PS3 \cite{ps3} shows decent performance under O0, with an F1 score of 0.83, but its performance degrades significantly under higher optimization levels (F1 score drops to 0.62, 0.64, and 0.64 for O1, O2, and O3, respectively). This degradation is due to the presence of too many patch-irrelevant features in its analysis, which introduces noise and leads to a high rate of missed patch detections.

Lares outperforms all baseline methods, showing robust performance across all optimization levels with an average F1 score of 0.77. Unlike BinXray \cite{BinXray}, Lares effectively handles cross-optimization scenarios by leveraging semantic features, which are less sensitive to syntactic changes introduced by optimization. Compared to PS3 \cite{ps3}, Lares reduces the influence of patch-irrelevant features through enhanced data flow, control flow, and macro analysis, improving recall even under high optimization levels. Finally, Lares surpasses Robin \cite{robin} by combining deterministic SMT-based verification with semantic reasoning, which enables it to handle complex cases where symbolic execution fails or extracted features are ambiguous. 

Specifically, we also evaluated 400 new cases, denoted as $Average_{new\_vul}$. These new vulnerabilities are consistent with the findings from conventional datasets, and Lares significantly improves over existing methods. This further demonstrates that Lares's improvement is not due to data contamination. This is consistent with our intuition, as Lares leverages the code analysis capabilities of LLM rather than its inherent understanding of vulnerabilities in the training set, and is therefore not affected by data contamination.


\begin{tcolorbox}[colback=gray!10,
                  colframe=black,
                  arc=1mm, auto outer arc,
                  boxrule=1.5pt,
                 ]

\textbf{Answering RQ1:} Lares can handle diverse optimization levels and significantly improve the precision and recall of existing methods.
\end{tcolorbox}

\subsection{Cross-architecture Task (RQ2)}

To evaluate the performance of Lares in the cross-architecture and cross-compiler patch presence testing task, we tested it on three architectures (x86, x64, ARM) and two compilers (GCC and Clang) with different optimization levels. Table~\ref{table_MAP2} summarizes the precision, recall, and F1 score of Lares under these configurations.

Existing methods such as BinXray \cite{BinXray}, PS3 \cite{ps3}, and Robin \cite{robin} rely heavily on architecture-specific or compiler-specific features, limiting their ability to handle cross-architecture or cross-compiler scenarios.  They are strongly related to instruction tokens, while different architectures have different instruction sets. 
Therefore, we only present the results of Lares. Unlike existing methods, Lares is explicitly designed to generalize across architectures and compilers.

Lares demonstrates consistent performance across the five settings. For x86 and x64 compiled with GCC, it achieves an average F1 score of 0.74, while for ARM-GCC, the score is also 0.71. Similarly, under Clang, the F1 scores range from 0.75 to 0.77. The slight variations across architectures are primarily due to differences in instruction sets and binary representations, but these have minimal impact on Lares' overall performance. Clang-based binaries show a slight improvement in F1 scores, possibly due to better preservation of semantic features during compilation. Similarly, IDA-Pro's decompilation for x86 architecture is more mature than ARM, so the result is slightly higher.



\begin{tcolorbox}[colback=gray!10,
                  colframe=black,
                  arc=1mm, auto outer arc,
                  boxrule=1.5pt,
                 ]

\textbf{Answering RQ2:} Lares outperforms existing methods by effectively handling the cross-architecture and cross-compiler patch presence testing task.
\end{tcolorbox}

\subsection{Ablation Evaluation (RQ3)}

We evaluated Lares with different LLMs and temperature parameters.  We have repeated every case 5 times when the temperature is not zero and observed no significant changes in the results. As shown in Table~\ref{tab:llm}, Claude-3.5-Sonet achieved the highest F1 score, followed by GPT-4o. The open-source model DeepSeek-Coder exhibited the highest both Recall but the lowest Precision. Claude-3.5 and GPT-4o significantly outperformed existing methods, while other models showed comparable performance. More advanced models, such as o1, were excluded due to cost constraints.

To evaluate the contribution of each module in Lares, we conducted an ablation study by progressively removing individual components. The results are summarized in Table~\ref{tab:freq4}, where the precision, recall, and F1 score of different variants of Lares are reported. The settings are same as Table~\ref{table_MAP}.


\begin{table}[t]
\centering

\renewcommand{\arraystretch}{1.2}
\caption{Performance of different LLMs across settings. Darker green indicates better performance}
\label{tab:llm}
\begin{tabular}{
    >{\raggedright\arraybackslash}p{0.1cm}  
    >{\raggedright\arraybackslash}p{2.5cm}!{\hspace{0.3pt}\vspace{0.3pt} }  
    *{4}{>{\centering\arraybackslash}p{0.8cm}!{\hspace{0.3pt}} } 
}
\multicolumn{2}{c}{} & 
\multicolumn{4}{c}{\textbf{Temperature}} \\
\addlinespace[3pt]
\multicolumn{2}{c}{\textbf{Metrics, Model} } & 
\cellcolor{lightgray}\textbf{0.0} & 
\cellcolor{lightgray}\textbf{0.5} & 
\cellcolor{lightgray}\textbf{0.7} & 
\cellcolor{lightgray}\textbf{1.0} \\
\addlinespace[3pt]
\multirow{5}{*}{\rotatebox[origin=c]{90}{\textbf{Precision}}} 
  & \cellcolor{lightgray} {\small  GPT-3.5-Turbo}
    &  0.514
    &  0.526
    &  0.543
    &  0.525 \\
  & \cellcolor{lightgray} {\small DeepSeek-Coder }
    &  0.511
    &  0.510
    &  0.515
    &  0.536 \\
  & \cellcolor{lightgray} {\small  GPT-4o} 
    &  0.586 
    &  0.598
    &  \cellcolor{lightgreen!50} 0.618 
    &  \cellcolor{lightgreen!50} 0.647  \\[2pt]
& \cellcolor{lightgray} {\small  Claude-3.5-Sonnet}
    & \cellcolor{lightgreen!50} 0.629
    & \cellcolor{lightgreen} 0.665
    & \cellcolor{green} 0.706
    & \cellcolor{green} 0.723 \\
\addlinespace[3pt]
\multirow{5}{*}{\rotatebox[origin=c]{90}{\textbf{Recall}}} 
  & \cellcolor{lightgray} {\small  GPT-3.5-Turbo}
    & \cellcolor{lightgreen} 0.814
    & \cellcolor{lightgreen} 0.826
    & \cellcolor{lightgreen} 0.829
    & \cellcolor{lightgreen!50} 0.792 \\
  & \cellcolor{lightgray} {\small DeepSeek-Coder }
    & \cellcolor{green} 0.983
    & \cellcolor{green} 0.934
    & \cellcolor{green} 0.951
    & \cellcolor{green} 0.968 \\
& \cellcolor{lightgray} {\small  GPT-4o} 
    &  \cellcolor{lightgreen} 0.824 
    &  \cellcolor{green} 0.923 
    &  \cellcolor{lightgreen} 0.885 
    &  \cellcolor{lightgreen} 0.872 \\[2pt]
  & \cellcolor{lightgray} {\small  Claude-3.5-Sonnet}
    &  0.673
    & \cellcolor{lightgreen} 0.802
    &  \cellcolor{lightgreen!50} 0.769
    & \cellcolor{lightgreen} 0.831 \\
\addlinespace[3pt]
\multirow{5}{*}{\rotatebox[origin=c]{90}{\textbf{F1 Score}}} 
  & \cellcolor{lightgray}{\small GPT-3.5-Turbo}
    & 0.631
    & 0.643
    & \cellcolor{lightgreen!50}0.656
    & 0.631 
      \\
  & \cellcolor{lightgray}{\small DeepSeek-Coder }
    & \cellcolor{lightgreen!50}0.673
    & \cellcolor{lightgreen!50}0.660
    & \cellcolor{lightgreen!50}0.668
    & \cellcolor{lightgreen!50}0.690 \\
& \cellcolor{lightgray}{\small GPT-4o} 
    & \cellcolor{lightgreen!50}0.685 
    & \cellcolor{lightgreen}0.726
    & \cellcolor{lightgreen}0.728
    & \cellcolor{lightgreen} 0.743  \\[2pt]
  & \cellcolor{lightgray}{\small Claude-3.5-Sonnet}
    & \cellcolor{lightgreen!50}0.651  
    & \cellcolor{lightgreen}0.727
    & \cellcolor{lightgreen}0.736
    & \cellcolor{green}0.773 \\
\addlinespace[3pt]
\end{tabular}

\end{table}

\begin{table}[t]
\renewcommand{\arraystretch}{1} 
  \centering
  \caption{Ablation study}
  \vspace{-1em}
  \label{tab:freq4}
  \begin{tabular}{cccc}
    \toprule
     Model & Precision & Recall &  F1\\
    \midrule
    $Lares_{-pe}$ & 0.651 & 0.720 &   0.683\\
    $Lares_{-pl}$ & 0.488 & 0.600 &   0.538\\
    $Lares_{-z3}$ &  0.673 & 0.780 & 0.714\\
    $Lares_{-pv}$ &  0.815 & 0.221 & 0.348\\
    $Lares$ & 0.725 &  0.825 & 0.772\\
 \bottomrule
\end{tabular}

\end{table}

Removing the patch enhancement module, $Lares_{-pe}$, degenerates into a common LLM-based function similarity matching method, which results in a precision of 0.651, recall of 0.720, and an F1 score of 0.683. This decline is due to the failure of the LLM to distinguish subtle differences in certain patches without the enriched semantic information provided by this module. 
When the patch localization module ($Lares_{-pl}$) is removed, and the LLM is directly tasked with detecting the presence of patches from pseudo code, the performance drops significantly, with an F1 score of 0.538. The absence of patch localization results in the LLM needing to process raw pseudo code with many missing variable names and structural ambiguities, which makes it difficult to reliably determine patch presence. 
This highlights the importance of the patch localization module in narrowing down the pseudo code to relevant slices, enabling the LLM to focus on smaller, semantically meaningful code regions.

Removing the Z3 solver ($Lares_{-z3}$) leads to a slight performance drop, with an F1 score of 0.714. Z3 is important and can improve F1-score from 71.4\% to 87.3\%. However, the F1-score for Lares is 77.3\%(not 87.3\%) because Z3 only handles 21\% testcases; the others, without equations for Z3 to prove semantic equivalence, are handled by LLM. Therefore, Z3 is necessary because when it works, the results will be highly confident. LLM is complementary to advancing scalability. In addition, the accuracy of LLM cannot be ignored. Although the LLM is less accurate than Z3, LLM-only performance still surpasses PS3 \cite{ps3} and Robin \cite{robin}.

Removing the LLM-based patch verification ($Lares_{-pv}$) and relying solely on Z3 for verification results in significant performance changes. While the precision of $Lares_{-pv}$ is high (0.815), the recall drops substantially to 0.221, as only 21\% of test cases can be successfully verified by the Z3 solver. This limitation stems from Z3's dependency on patches that can be fully modeled using logical constraints, whereas many real-world patches lack unique semantic equations suitable for Z3's processing. These results highlight that, although Z3 achieves high precision, its applicability is limited, requiring LLM to deal with more cases.

\begin{figure}
\centering

    \includegraphics[width=0.9\linewidth]{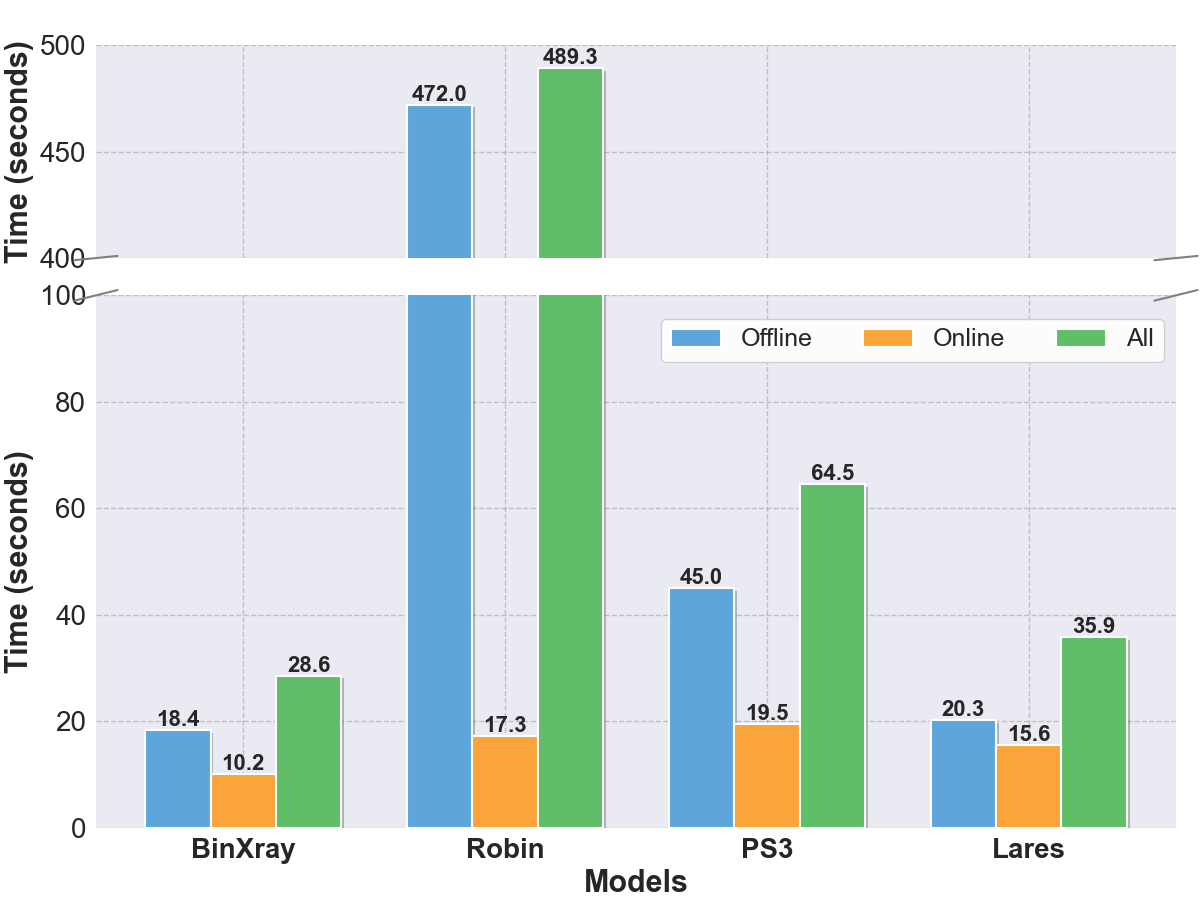}

\caption{Timecost Evaluation (seconds).}
\label{fig:timecost}
\vspace{-1.0em}
\end{figure}

\begin{tcolorbox}[colback=gray!10,
                  colframe=black,
                  arc=1mm, auto outer arc,
                  boxrule=1.5pt,
                 ]

\textbf{Answering RQ3:} Leveraging LLM, Lares significantly surpasses existing methods. Each component contributes critically to its overall performance.
\end{tcolorbox}

\subsection{Efficiency (RQ4)}

To evaluate the efficiency of Lares, we compared its runtime with existing approaches. The runtime is divided into two phases: \emph{OffLine}, which includes offline preprocessing tasks such as reverse engineering and feature extraction for patches, and \emph{OnLine}, which refers to the patch presence testing time for each input binary. The results are summarized in Figure~\ref{fig:timecost}.

BinXray \cite{BinXray} has the shortest runtime, with a total time of 28.6 seconds. This is because it uses only simple syntactic features. Robin \cite{robin} is the most time-consuming among all methods, with a total runtime of 489.31 seconds. The primary bottleneck is the \emph{OffLine} phase, which takes 472 seconds due to the use of symbolic execution for feature extraction. PS3 \cite{ps3} improves efficiency compared to Robin \cite{robin} by simplifying symbolic execution into symbolic simulation, significantly reducing the time required in the \emph{OffLine} phase to 45 seconds. 

Lares achieves a total runtime of 35.9 seconds, which is significantly faster than Robin \cite{robin} and PS3 \cite{ps3} and close to the lightweight BinXray \cite{BinXray}. Its \emph{OffLine} phase takes only 20.3 seconds, as it avoids the computational overhead of symbolic execution by leveraging semantic analysis based on a large language model (LLM). The \emph{OnLine} phase (15.6 seconds) is also reasonably fast, enabling efficient patch detection without compromising accuracy.



The runtime comparison in Figure~\ref{fig:timecost} does not include the time required for compiling binaries. BinXray \cite{BinXray}, Robin \cite{robin} and PS3 \cite{ps3} require a compilation process to construct patch features, which adds significant overhead in real-world use.

\begin{tcolorbox}[colback=gray!10,
                  colframe=black,
                  arc=1mm, auto outer arc,
                  boxrule=1.5pt,
                 ]

\textbf{Answering RQ4:} Lares outperforms existing methods and particularly suitable for scenarios where speed and adaptability are crucial.
\end{tcolorbox}

\subsection{Failure Cases Analysis}
We investigate various failure cases in this experiment and identify several potential causes behind these inaccuracies:

\begin{itemize}
    \item \textbf{Repeated patch code}: The statements added by patches that are also in the vulnerable function, but in different locations to interfere with judgment. For example, the if statement ``if ($\mathtt{s \rightarrow session \rightarrow sess\_cert == NULL}$)" in CVE-2014-3510 and the code in the body also appear elsewhere in the vulnerable function. This is a relatively common assertion. Therefore, Lares also recognizes this statement in the vulnerable function and mistakenly identifies the vulnerable function as patched. This situation is rare in our dataset. In the future, researchers can optimize the Patch Enhancement module to avoid this failure by adding more context.
    \item \textbf{Stealthy patches}: Stealthy patches often introduce only minor changes. Variations from different compiler environments further obscure these differences, making accurate judgment difficult. In fact, these cases are also difficult for humans. Specifically, the LLM shows high precision for add/delete-type patches but occasionally fails on edit-type patches by missing subtle differences or misidentifying changes.
    \item \textbf{The hallucinations of LLM}: Hallucination is one of the recognized problems in LLM.  Hallucinations can degrade performance by causing missed patches or incorrect judgments, especially in logical comparisons and complex vulnerabilities. For instance, in CVE-2013-6449 the condition changed from “$>=$” to “$==$” post-patch, yet the LLM sometimes hallucinated a “$==$” and falsely deemed the vulnerability patched. Future work will mitigate hallucinations via iterative verification and building more complex multi-agent workflows.
\end{itemize}

In summary, Lares demonstrated that the agent built by combining LLM and Z3 has considerable potential in the patch existence testing task. Lares cannot achieve perfect judgment in some special cases due to the presence of hallucinations. However, we can try to address these issues in the future by improving the agent or designing workflows based on vulnerability types.

%% file: tex/discuss.tex
\section{Discussion}


Patch existence testing, which involves detecting fine-grained patch semantics, remains a challenging task with significant room for improvement in existing methods, including Lares. 
Lares offers interpretable outputs, such as equations generated by Z3 or reasoning provided by LLMs, which can serve as valuable references for humans, even in cases of uncertain results.

In this work, we employed a zero-shot commercial LLM for patch presence testing and achieved promising results. However, this approach incurs a certain API usage cost. In the future, by fine-tuning open-source LLMs on patch-related data, it is possible to further enhance detection performance while reducing the cost of deployment. Besides, Sometimes Lares will return a result that is not in JSON format because of the randomness of LLM, and we need to ask again. 
 
Our dataset includes a diverse set of vulnerable and patched functions, comprising different types of projects. This dataset is consistent with previous work \cite{ps3} but goes beyond by generating significantly more test cases, with a total of 4480 samples. The larger dataset ensures a more thorough and reliable evaluation of Lares' performance, covering a wide range of real-world scenarios and challenges.

%% file: tex/related.tex
\section{Related Work}

\subsection{Vulnerability Detection}

Binary code similarity detection plays a crucial \cite{Code_Similarity_survey1, Code_Similarity_survey2} role in applications like malware analysis \cite{BinSim}, vulnerability detection \cite{QF_bugsearch, IR-Cross_arch_bug_search}, and software reuse identification \cite{ISRD, LibDB, LibDI, LibDX}. This work focuses on detecting 1-day vulnerabilities caused by binary code reuse \cite{Vulhawk, jTrans, OSCAR, Asteria-pro}. 
The main challenge is to overcome the differences caused by different compilation environments and identify semantic equivalence.

Recent works leverage Natural Language Processing (NLP) techniques and structural analysis for binary similarity detection. NLP-based methods, such as SAFE \cite{SAFE}, Trex \cite{Trex}, and CEBin \cite{cebin}, treat code as natural language and embed semantics for matching. Structural approaches like Gemini \cite{Gemini, libvdiff}, ASTeria \cite{Asteria}, and HermesSim \cite{HermesSim} exploit control flow graphs (CFG), abstract syntax trees (AST), or custom semantic graphs for similarity detection.

Large language models (LLMs) have recently emerged as a new paradigm in this field. Methods like CLAP \cite{clap} and SCALE \cite{scale} use LLMs to enhance similarity detection.

Despite these advances, current methods struggle to determine if detected vulnerabilities have been patched. To address this, our work targets patch presence testing, which complements vulnerability detection by verifying whether a vulnerability has been mitigated, offering a more comprehensive assessment of software security.

\subsection{Vulnerability Verification}  
To address false positives in vulnerability detection, verification methods are employed, including dynamic and static approaches. 

Dynamic methods, such as Directed Grey-box Fuzzing (DGF) \cite{dafl, kafl, aflplus ,selectfuzz, beacon}, generate proofs of concept (PoC) by fuzzing target binaries \cite{octopocs}. AFLGo \cite{aflgo} is a classic DGF approach. Subsequent methods like Hawkeye \cite{hawkeye}, WindRanger \cite{WindRanger}, and TransferFuzz \cite{TransferFuzz} build upon AFLGo, enabling direct vulnerability triggering. However, these methods are time-consuming and limited to a small subset of memory vulnerabilities.

To overcome these limitations, static methods like patch existence testing have been proposed. These approaches analyze patch modification statements to detect whether a function remains vulnerable or has been patched. 
Some researchers \cite{patch_survey1, patch_survey2} have analyzed the evolution of vulnerability patches and classified various security patches. Patch existence testing methods \cite{AndroidPatch} for Java, Android, or C/C++ source code are relatively mature. However, patch existence testing for C/C++ binary targets brings unique challenges due to different compilation environments.
BinXray \cite{BinXray} uses lightweight syntactic features. Robin \cite{robin} and PS3 \cite{ps3} leverage symbolic execution and simulation for higher accuracy but face scalability challenges. Specialized kernel-focused methods, such as Fiber \cite{Fiber} and PDiff \cite{Pdiff}, have also been developed.

Existing approaches often trade off accuracy for usability or vice versa. We propose \textit{Lares}, a lightweight method that achieves both high usability and accuracy while uniquely supporting cross-optimization, cross-architecture, and cross-compiler scenarios.

%% file: main.bbl
\begin{thebibliography}{10}

\bibitem{FirmSec}
Binbin Zhao, Shouling Ji, Jiacheng Xu, Yuan Tian, Qiuyang Wei, Qinying Wang, Chenyang Lyu, Xuhong Zhang, Changting Lin, Jingzheng Wu, et~al.
\newblock A large-scale empirical analysis of the vulnerabilities introduced by third-party components in iot firmware.
\newblock In {\em Proceedings of the 31st ACM SIGSOFT International Symposium on Software Testing and Analysis}, pages 442--454, 2022.

\bibitem{Blackduck}
Synopsys.
\newblock Synopsys 2024 open source security and risk analysis report., 2024.
\newblock \url{https://www.blackduck.com/resources/analyst-reports/open-source-security-risk-analysis.html}.

\bibitem{Gemini}
Xiaojun Xu, Chang Liu, Qian Feng, Heng Yin, Le~Song, and Dawn Song.
\newblock Neural network-based graph embedding for cross-platform binary code similarity detection.
\newblock In {\em Proceedings of the 2017 ACM SIGSAC Conference on Computer and Communications Security}, pages 363--376, 2017.

\bibitem{jTrans}
Hao Wang, Wenjie Qu, Gilad Katz, Wenyu Zhu, Zeyu Gao, Han Qiu, Jianwei Zhuge, and Chao Zhang.
\newblock Jtrans: Jump-aware transformer for binary code similarity detection.
\newblock In {\em Proceedings of the 31st ACM SIGSOFT International Symposium on Software Testing and Analysis}, pages 1--13, 2022.

\bibitem{Vulhawk}
Zhenhao Luo, Pengfei Wang, Baosheng Wang, Yong Tang, Wei Xie, Xu~Zhou, Danjun Liu, and Kai Lu.
\newblock Vulhawk: Cross-architecture vulnerability detection with entropy-based binary code search.
\newblock In {\em NDSS}, 2023.

\bibitem{LibDB}
Wei Tang, Yanlin Wang, Hongyu Zhang, Shi Han, Ping Luo, and Dongmei Zhang.
\newblock Libdb: an effective and efficient framework for detecting third-party libraries in binaries.
\newblock In {\em Proceedings of the 19th International Conference on Mining Software Repositories}, pages 423--434, 2022.

\bibitem{LibAM}
Siyuan Li, Yongpan Wang, Chaopeng Dong, Shouguo Yang, Hong Li, Hao Sun, Zhe Lang, Zuxin Chen, Weijie Wang, Hongsong Zhu, et~al.
\newblock Libam: An area matching framework for detecting third-party libraries in binaries.
\newblock {\em ACM Transactions on Software Engineering and Methodology}, 33(2):1--35, 2023.

\bibitem{binaryai}
Ling Jiang, Junwen An, Huihui Huang, Qiyi Tang, Sen Nie, Shi Wu, and Yuqun Zhang.
\newblock Binaryai: Binary software composition analysis via intelligent binary source code matching.
\newblock In {\em Proceedings of the IEEE/ACM 46th International Conference on Software Engineering}, pages 1--13, 2024.

\bibitem{robin}
Shouguo Yang, Zhengzi Xu, Yang Xiao, Zhe Lang, Wei Tang, Yang Liu, Zhiqiang Shi, Hong Li, and Limin Sun.
\newblock Towards practical binary code similarity detection: Vulnerability verification via patch semantic analysis.
\newblock {\em ACM Trans. Softw. Eng. Methodol.}, 32(6), sep 2023.

\bibitem{Fiber}
Hang Zhang and Zhiyun Qian.
\newblock Precise and accurate patch presence test for binaries.
\newblock In {\em 27th USENIX Security Symposium (USENIX Security 18)}, pages 887--902, 2018.

\bibitem{BinXray}
Yifei Xu, Zhengzi Xu, Bihuan Chen, Fu~Song, Yang Liu, and Ting Liu.
\newblock Patch based vulnerability matching for binary programs.
\newblock In {\em Proceedings of the 29th ACM SIGSOFT International Symposium on Software Testing and Analysis}, pages 376--387, 2020.

\bibitem{patchdiscovery}
Xi~Xu, Qinghua Zheng, Zheng Yan, Ming Fan, Ang Jia, Zhaohui Zhou, Haijun Wang, and Ting Liu.
\newblock Patchdiscovery: Patch presence test for identifying binary vulnerabilities based on key basic blocks.
\newblock {\em IEEE Transactions on Software Engineering}, 49(12):5279--5294, 2023.

\bibitem{Pdiff}
Zheyue Jiang, Yuan Zhang, Jun Xu, Qi~Wen, Zhenghe Wang, Xiaohan Zhang, Xinyu Xing, Min Yang, and Zhemin Yang.
\newblock Pdiff: Semantic-based patch presence testing for downstream kernels.
\newblock In {\em Proceedings of the 2020 ACM SIGSAC Conference on Computer and Communications Security}, pages 1149--1163, 2020.

\bibitem{ps3}
Qi~Zhan, Xing Hu, Zhiyang Li, Xin Xia, David Lo, and Shanping Li.
\newblock Ps3: Precise patch presence test based on semantic symbolic signature.
\newblock In {\em Proceedings of the IEEE/ACM 46th International Conference on Software Engineering}, pages 1--12, 2024.

\bibitem{react}
Qi~Zhan, Xing Hu, Xin Xia, and Shanping Li.
\newblock React: Ir-level patch presence test for binary.
\newblock In {\em Proceedings of the 39th IEEE/ACM International Conference on Automated Software Engineering}, pages 381--392, 2024.

\bibitem{llm4decompile}
Hanzhuo Tan, Qi~Luo, Jing Li, and Yuqun Zhang.
\newblock Llm4decompile: Decompiling binary code with large language models.
\newblock {\em arXiv preprint arXiv:2403.05286}, 2024.

\bibitem{degpt}
Peiwei Hu, Ruigang Liang, and Kai Chen.
\newblock Degpt: Optimizing decompiler output with llm.
\newblock In {\em Proceedings 2024 Network and Distributed System Security Symposium (2024). https://api. semanticscholar. org/CorpusID}, volume 267622140, 2024.

\bibitem{B2SFinder}
Zimu Yuan, Muyue Feng, Feng Li, Gu~Ban, Yang Xiao, Shiyang Wang, Qian Tang, He~Su, Chendong Yu, Jiahuan Xu, et~al.
\newblock B2sfinder: detecting open-source software reuse in cots software.
\newblock In {\em 2019 34th IEEE/ACM International Conference on Automated Software Engineering (ASE)}, pages 1038--1049. IEEE, 2019.

\bibitem{ISRD}
Xi~Xu, Qinghua Zheng, Zheng Yan, Ming Fan, Ang Jia, and Ting Liu.
\newblock Interpretation-enabled software reuse detection based on a multi-level birthmark model.
\newblock In {\em 2021 IEEE/ACM 43rd International Conference on Software Engineering (ICSE)}, pages 873--884. IEEE, 2021.

\bibitem{MVP}
Yang Xiao, Bihuan Chen, Chendong Yu, Zhengzi Xu, Zimu Yuan, Feng Li, Binghong Liu, Yang Liu, Wei Huo, Wei Zou, et~al.
\newblock $\{$MVP$\}$: Detecting vulnerabilities using $\{$Patch-Enhanced$\}$ vulnerability signatures.
\newblock In {\em 29th USENIX Security Symposium (USENIX Security 20)}, pages 1165--1182, 2020.

\bibitem{joern}
Joern., 2014.
\newblock \url{https://joern.io/}.

\bibitem{IDA_Pro}
Ida pro., 2023.
\newblock \url{https://hex-rays.com/IDA-pro/}.

\bibitem{treesitter}
tree-sitter., 2017.
\newblock \url{https://github.com/tree-sitter/tree-sitter}.

\bibitem{z3}
Z3 prover., 2023.
\newblock \url{https://github.com/Z3Prover/z3}.

\bibitem{claude}
Claude., 2023.
\newblock \url{https://claude.ai/}.

\bibitem{Code_Similarity_survey1}
Saed Alrabaee, Mourad Debbabi, and Lingyu Wang.
\newblock A survey of binary code fingerprinting approaches: Taxonomy, methodologies, and features.
\newblock {\em ACM Computing Surveys (CSUR)}, 55(1):1--41, 2022.

\bibitem{Code_Similarity_survey2}
Irfan~Ul Haq and Juan Caballero.
\newblock A survey of binary code similarity.
\newblock {\em ACM Computing Surveys (CSUR)}, 54(3):1--38, 2021.

\bibitem{BinSim}
Jiang Ming, Dongpeng Xu, Yufei Jiang, and Dinghao Wu.
\newblock $\{$BinSim$\}$: Trace-based semantic binary diffing via system call sliced segment equivalence checking.
\newblock In {\em 26th USENIX Security Symposium (USENIX Security 17)}, pages 253--270, 2017.

\bibitem{QF_bugsearch}
Qian Feng, Minghua Wang, Mu~Zhang, Rundong Zhou, Andrew Henderson, and Heng Yin.
\newblock Extracting conditional formulas for cross-platform bug search.
\newblock In {\em Proceedings of the 2017 ACM on Asia Conference on Computer and Communications Security}, pages 346--359, 2017.

\bibitem{IR-Cross_arch_bug_search}
Jannik Pewny, Behrad Garmany, Robert Gawlik, Christian Rossow, and Thorsten Holz.
\newblock Cross-architecture bug search in binary executables.
\newblock In {\em 2015 IEEE Symposium on Security and Privacy}, pages 709--724. IEEE, 2015.

\bibitem{LibDI}
Siyuan Li, Chaopeng Dong, Yongpan Wang, Wenming Liu, Weijie Wang, Hong Li, Hongsong Zhu, and Limin Sun.
\newblock Libdi: A direction identification framework for detecting complex reuse relationships in binaries.
\newblock In {\em MILCOM 2023-2023 IEEE Military Communications Conference (MILCOM)}, pages 741--746. IEEE, 2023.

\bibitem{LibDX}
Wei Tang, Ping Luo, Jialiang Fu, and Dan Zhang.
\newblock Libdx: A cross-platform and accurate system to detect third-party libraries in binary code.
\newblock In {\em 2020 IEEE 27th International Conference on Software Analysis, Evolution and Reengineering (SANER)}, pages 104--115. IEEE, 2020.

\bibitem{OSCAR}
Dinglan Peng, Shuxin Zheng, Yatao Li, Guolin Ke, Di~He, and Tie-Yan Liu.
\newblock How could neural networks understand programs?
\newblock In {\em International Conference on Machine Learning}, pages 8476--8486. PMLR, 2021.

\bibitem{Asteria-pro}
Shouguo Yang, Chaopeng Dong, Yang Xiao, Yiran Cheng, Zhiqiang Shi, Zhi Li, and Limin Sun.
\newblock Asteria-pro: Enhancing deep-learning based binary code similarity detection by incorporating domain knowledge.
\newblock {\em arXiv preprint arXiv:2301.00511}, 2023.

\bibitem{SAFE}
Luca Massarelli, Giuseppe Antonio~Di Luna, Fabio Petroni, Roberto Baldoni, and Leonardo Querzoni.
\newblock Safe: Self-attentive function embeddings for binary similarity.
\newblock In {\em International Conference on Detection of Intrusions and Malware, and Vulnerability Assessment}, pages 309--329. Springer, 2019.

\bibitem{Trex}
Kexin Pei, Zhou Xuan, Junfeng Yang, Suman Jana, and Baishakhi Ray.
\newblock Trex: Learning execution semantics from micro-traces for binary similarity.
\newblock {\em arXiv preprint arXiv:2012.08680}, 2020.

\bibitem{cebin}
Hao Wang, Zeyu Gao, Chao Zhang, Mingyang Sun, Yuchen Zhou, Han Qiu, and Xi~Xiao.
\newblock Cebin: A cost-effective framework for large-scale binary code similarity detection.
\newblock In {\em Proceedings of the 33rd ACM SIGSOFT International Symposium on Software Testing and Analysis}, pages 149--161, 2024.

\bibitem{libvdiff}
Chaopeng Dong, Siyuan Li, Shouguo Yang, Yang Xiao, Yongpan Wang, Hong Li, Zhi Li, and Limin Sun.
\newblock Libvdiff: Library version difference guided oss version identification in binaries.
\newblock In {\em Proceedings of the 46th IEEE/ACM International Conference on Software Engineering}, ICSE '24, New York, NY, USA, 2024. Association for Computing Machinery.

\bibitem{Asteria}
Shouguo Yang, Long Cheng, Yicheng Zeng, Zhe Lang, Hongsong Zhu, and Zhiqiang Shi.
\newblock Asteria: Deep learning-based ast-encoding for cross-platform binary code similarity detection.
\newblock In {\em 2021 51st Annual IEEE/IFIP International Conference on Dependable Systems and Networks (DSN)}, pages 224--236. IEEE, 2021.

\bibitem{HermesSim}
Haojie He, Xingwei Lin, Ziang Weng, Ruijie Zhao, Shuitao Gan, Libo Chen, Yuede Ji, Jiashui Wang, and Zhi Xue.
\newblock Code is not natural language: Unlock the power of semantics-oriented graph representation for binary code similarity detection.
\newblock In {\em 33rd USENIX Security Symposium (USENIX Security 24), PHILADELPHIA, PA}, 2024.

\bibitem{clap}
Hao Wang, Zeyu Gao, Chao Zhang, Zihan Sha, Mingyang Sun, Yuchen Zhou, Wenyu Zhu, Wenju Sun, Han Qiu, and Xi~Xiao.
\newblock Clap: Learning transferable binary code representations with natural language supervision.
\newblock In {\em Proceedings of the 33rd ACM SIGSOFT International Symposium on Software Testing and Analysis}, pages 503--515, 2024.

\bibitem{scale}
Xin-Cheng Wen, Cuiyun Gao, Shuzheng Gao, Yang Xiao, and Michael~R Lyu.
\newblock Scale: Constructing structured natural language comment trees for software vulnerability detection.
\newblock In {\em Proceedings of the 33rd ACM SIGSOFT International Symposium on Software Testing and Analysis}, pages 235--247, 2024.

\bibitem{dafl}
Tae~Eun Kim, Jaeseung Choi, Kihong Heo, and Sang~Kil Cha.
\newblock $\{$DAFL$\}$: Directed grey-box fuzzing guided by data dependency.
\newblock In {\em 32nd USENIX Security Symposium (USENIX Security 23)}, pages 4931--4948, 2023.

\bibitem{kafl}
Sergej Schumilo, Cornelius Aschermann, Robert Gawlik, Sebastian Schinzel, and Thorsten Holz.
\newblock $\{$kAFL$\}$:$\{$Hardware-Assisted$\}$ feedback fuzzing for $\{$OS$\}$ kernels.
\newblock In {\em 26th USENIX security symposium (USENIX Security 17)}, pages 167--182, 2017.

\bibitem{aflplus}
Andrea Fioraldi, Dominik Maier, Heiko Ei{\ss}feldt, and Marc Heuse.
\newblock $\{$AFL++$\}$: Combining incremental steps of fuzzing research.
\newblock In {\em 14th USENIX Workshop on Offensive Technologies (WOOT 20)}, 2020.

\bibitem{selectfuzz}
Changhua Luo, Wei Meng, and Penghui Li.
\newblock Selectfuzz: Efficient directed fuzzing with selective path exploration.
\newblock In {\em 2023 IEEE Symposium on Security and Privacy (SP)}, pages 2693--2707, 2023.

\bibitem{beacon}
Heqing Huang, Yiyuan Guo, Qingkai Shi, Peisen Yao, Rongxin Wu, and Charles Zhang.
\newblock Beacon: Directed grey-box fuzzing with provable path pruning.
\newblock In {\em 2022 IEEE Symposium on Security and Privacy (SP)}, pages 36--50. IEEE, 2022.

\bibitem{octopocs}
Seongkyeong Kwon, Seunghoon Woo, Gangmo Seong, and Heejo Lee.
\newblock Octopocs: automatic verification of propagated vulnerable code using reformed proofs of concept.
\newblock In {\em 2021 51st Annual IEEE/IFIP International Conference on Dependable Systems and Networks (DSN)}, pages 174--185. IEEE, 2021.

\bibitem{aflgo}
Marcel B{\"o}hme, Van-Thuan Pham, Manh-Dung Nguyen, and Abhik Roychoudhury.
\newblock Directed greybox fuzzing.
\newblock In {\em Proceedings of the 2017 ACM SIGSAC conference on computer and communications security}, pages 2329--2344, 2017.

\bibitem{hawkeye}
Hongxu Chen, Yinxing Xue, Yuekang Li, Bihuan Chen, Xiaofei Xie, Xiuheng Wu, and Yang Liu.
\newblock Hawkeye: Towards a desired directed grey-box fuzzer.
\newblock In {\em Proceedings of the 2018 ACM SIGSAC conference on computer and communications security}, pages 2095--2108, 2018.

\bibitem{WindRanger}
Zhengjie Du, Yuekang Li, Yang Liu, and Bing Mao.
\newblock Windranger: a directed greybox fuzzer driven by deviation basic blocks.
\newblock In {\em Proceedings of the 44th International Conference on Software Engineering}, pages 2440--2451, 2022.

\bibitem{TransferFuzz}
Siyuan Li, Yuekang Li, Zuxin Chen, Chaopeng Dong, Yongpan Wang, Hong Li, Yongle Chen, and Hongsong Zhu.
\newblock Transferfuzz: Fuzzing with historical trace for verifying propagated vulnerability code.
\newblock {\em arXiv preprint arXiv:2411.18347}, 2024.

\bibitem{patch_survey1}
Zhiwei Fei, Jidong Ge, Chuanyi Li, Tianqi Wang, Yuning Li, Haodong Zhang, LiGuo Huang, and Bin Luo.
\newblock Patch correctness assessment: A survey.
\newblock {\em ACM Transactions on Software Engineering and Methodology}, 34(2):1--50, 2025.

\bibitem{patch_survey2}
Ruyan Lin, Yulong Fu, Wei Yi, Jincheng Yang, Jin Cao, Zhiqiang Dong, Fei Xie, and Hui Li.
\newblock Vulnerabilities and security patches detection in oss: a survey.
\newblock {\em ACM Computing Surveys}, 57(1):1--37, 2024.

\bibitem{AndroidPatch}
Zifan Xie, Ming Wen, Haoxiang Jia, Xiaochen Guo, Xiaotong Huang, Deqing Zou, and Hai Jin.
\newblock Precise and efficient patch presence test for android applications against code obfuscation.
\newblock In {\em Proceedings of the 32nd ACM SIGSOFT International Symposium on Software Testing and Analysis}, pages 347--359, 2023.

\end{thebibliography}
